\documentclass{article}
\usepackage{spconf,amsmath,graphicx,hyperref}
\usepackage[normalem]{ulem}
\usepackage[T1]{fontenc}
\usepackage{booktabs}
\usepackage{multirow}
\usepackage{pifont}
\usepackage{graphicx}
\usepackage{amssymb}
\makeatletter
\newcommand{\rmnum}[1]{\romannumeral #1}
\newcommand{\Rmnum}[1]{\expandafter\@slowromancap\romannumeral #1@}
\makeatother
\usepackage{stfloats} 
\usepackage{enumitem}

\usepackage[table]{xcolor}
\usepackage{makecell}
\usepackage{siunitx}
\definecolor{myblue}{RGB}{233, 241, 249}
\definecolor{mygray}{RGB}{99, 110, 114}
\definecolor{myred}{RGB}{255, 118, 117}
\definecolor{myyellow}{RGB}{255, 234, 167}
\definecolor{mygreen}{RGB}{216, 226, 204}
\definecolor{mypurple}{RGB}{162, 155, 254}
\definecolor{mybrown}{RGB}{215, 190, 154}
\definecolor{myorange}{RGB}{255, 220, 190}

\title{FINRS: A Risk-Sensitive Trading Framework for Real Financial Markets}
\twoauthors
 {Bijia Liu}
 {Dongbei University of Finance and Economics}
 {Ronghao Dang\sthanks{Corresponding author}}
{Alibaba DAMO Academy}

\begin{document}
%
\maketitle

\begin{abstract}

Large language models (LLMs) have shown strong reasoning capabilities and are increasingly explored for financial trading. Existing LLM-based trading agents, however, largely focus on single-step prediction and lack integrated mechanisms for risk management, which reduces their effectiveness in volatile markets. We introduce FinRS, a risk-sensitive trading framework that combines hierarchical market analysis, dual-decision agents, and multi-timescale reward reflection to align trading actions with both return objectives and downside risk constraints. Experiments on multiple stocks and market conditions show that FinRS achieves superior profitability and stability compared to state-of-the-art methods.

\keywords{Trading Agent, Risk-Sensitive, Real Markets}
\end{abstract}

\section{Introduction}

In recent years, large language models (LLMs)~\cite{llm, deepseek-v3} have demonstrated significant potential in financial trading. However, most existing methods focus on single-step predictions and overlook risks in multi-stage sequential trading. Existing works offer only surface-level remedies: FINMEM~\cite{finmem} adopts configuration-based risk preference switching; FinAgent~\cite{finagent} relies on risk-adjusted indicators for ex-post evaluation; FinCon~\cite{fincon} applies heuristic drawdown-control rules. While these designs offer a degree of mitigation for trading risks, they remain ad hoc and fail to capture the inherent risks of sequential trading. This limitation stems from several fundamental issues. First, task modeling is overly simplistic: most methods reduce the problem to a single-step classification, neglecting position sizing and temporal continuity. Second, risk perception is insufficient: feedback often targets returns while ignoring drawdowns~\cite{mdd} and exposure, rendering strategies fragile under extreme market conditions. Third, information processing is shallow: decision~\cite{llm_decision} chains typically follow a one-shot "collect–analyze–predict" pattern, lacking hierarchical filtering and causal focus, which makes them vulnerable to noise. Fourth, existing feedback schemes typically address risk in a reactive or rule-based manner (e.g., prompt-based adjustments, numerical alarms) rather than integrating risk-adjusted performance across multiple time scales.


To address these issues, we propose \textbf{FinRS}: a risk-sensitive LLM trading framework for multi-stage decision-making. Its key designs include: (\rmnum{1}) dynamic position sizing modeling to support more flexible risk exposure control; (\rmnum{2}) injecting risk-sensitive reasoning into the reasoning chain~\cite{llm_reasoning} so that decisions not only pursue returns but also provide insights into future stock trends; (\rmnum{3}) layered information filtering to gradually distill effective signals from reports and news; (\rmnum{4}) a hierarchical feedback mechanism that combines short-term returns with long-term robustness goals to guide trading consistency across time scales.






\section{Related Works}

Recent advances in large language models (LLMs) have enabled powerful reasoning and have been applied to financial trading agents~\cite{agent}. Prior work progressed from single-framework approaches (FINMEM~\cite{finmem}) to multimodal fusion (FinAgent~\cite{finagent}, FinVision~\cite{finvision}). Later, multi-agent systems with contextual reflection (FINCON~\cite{fincon}, TradingAgents~\cite{tradingagents}) further enhanced the cognitive and decision-making capabilities of financial agents. However, these systems address risk only superficially: FINMEM relies on simple preference switching, FinAgent uses ex-post risk metrics, and FINCON applies heuristic drawdown rules~\cite{risk_control}. Despite their reasoning strengths, existing LLM agents lack systematic risk modeling and control mechanisms, which limits their effectiveness in real-world trading scenarios.

\section{Architecture of FinRS} 

\begin{figure*}[htbp]
\centering
\includegraphics[width=\textwidth]{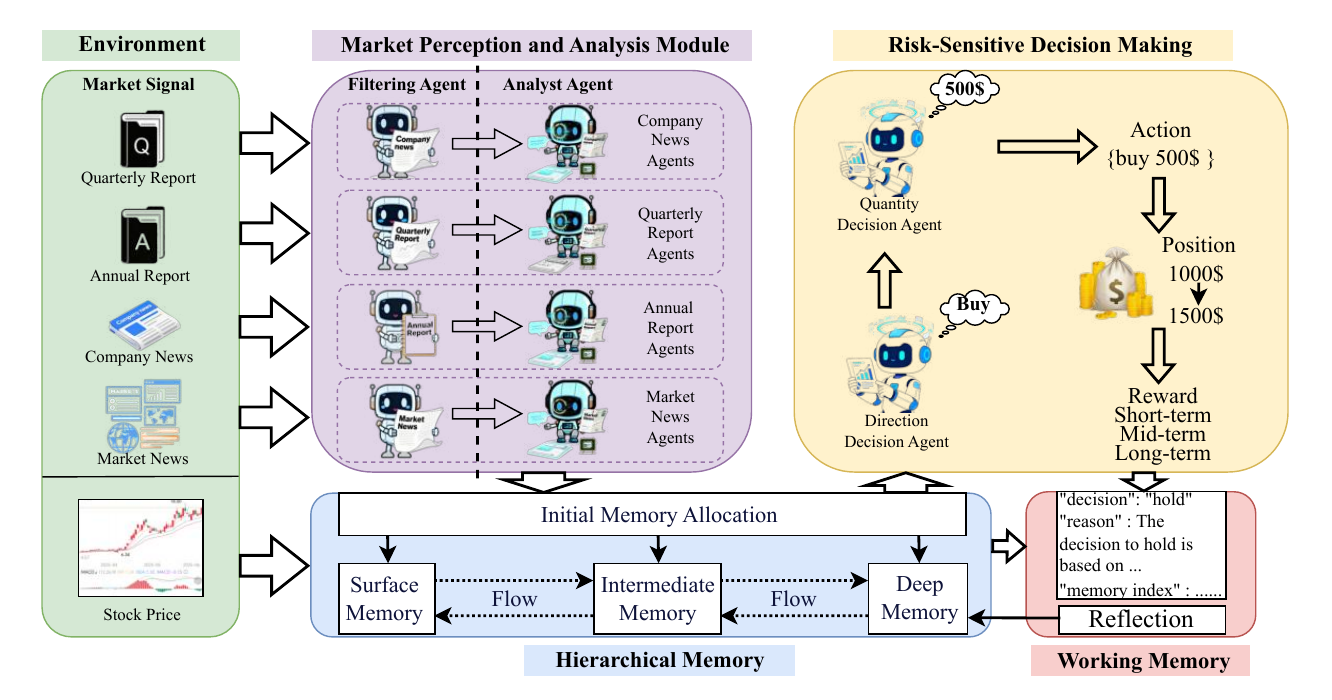}
\caption{\textbf{Architectural Details of FinRS:} A risk-sensitive design combining perception, memory allocation, dual decision agents, and reflective updates via multi-timescale rewards.}
\label{fig:archi}
\end{figure*}

To address these existing limitations, we propose a systematic risk-control framework that embeds risk-awareness directly into the decision-making process of LLM-based trading agents, thereby aligning them more closely with real-world financial practice. Specifically, the architecture of FinRS (Fig.~\ref{fig:archi}) is organized into three core modules: market observation and analysis, risk-sensitive decision making, and multi-scale reward reflection. We next elaborate on the role of each module and how they interact.




\subsection{Market Perception and Analysis Module}

Inspired by institutional workflows, this module (highlighted in purple in Fig.~\ref{fig:archi}) deploys groups of analyst agents, each focusing on a specific market signal such as news, financial statements, or macro narratives. Raw data first passes through a filtering stage, where agents assess relevance and potential price impact, and annotate reasons for inclusion for traceability. The filtered signals are then passed to their corresponding analyst agents, which operate under domain-specific prompt designs. 

In our experiments, we observed that relying solely on sentiment analysis often led to systematic biases, particularly during periods of high market uncertainty. To address this, we explicitly incorporate risk cues into the prompts, such as indicators of potential downside or signals of market volatility, guiding the agents to focus on risk-relevant factors while avoiding misleading signals.

By adopting this “pre-screening + specialized analysis” strategy, FinRS not only reduces noise but also highlights risk-relevant market factors, ensuring that trading decisions rest on more reliable evidence. 

Outputs are stored in hierarchical memory (highlighted in blue in Fig.~\ref{fig:archi}): stable signals (e.g., annual reports) in deep layers, volatile ones (e.g., breaking news) in shallow layers. The hierarchy is dynamic—profitable or protective signals are promoted deeper, while misleading ones are weakened or discarded. This adaptive design mirrors human trading experience, gradually enhancing the agent’s risk awareness.


\subsection{Risk-Sensitive Decision Making}
The risk-sensitive trading decision module (highlighted in yellow in Fig.~\ref{fig:archi}) is responsible for generating specific trading instructions in dynamic markets. Unlike previous research that only predicted direction, FinRS explicitly models position sizing in this module to measure an account's risk exposure, elevating "prediction" to "decision-making." This module continues FINMEM's "structured memory + decision-making agent" paradigm: the multi-source data representations generated by the market observation and analysis module are retrieved and provided to the decision-making agent, allowing it to make decisions based not on a single signal but on a combination of semantic information, historical experience, and current positions.

In this module, we set up two core agents: a direction decision agent, a quantity and risk agent. The direction agent is responsible for deciding the specific trading action—buy, sell, or hold—based on retrieved experience, recent market dynamics, and current position status. To enhance its robustness in highly volatile markets, FinRS integrates 1-day, 7-day, and 30-day price differences as momentum features, the model understands the tension between short-term volatility and medium- to long-term trends, thereby avoiding overreaction to short-term anomalies.

Once the directional decision is made, the quantity and risk agents are activated. Unlike traditional methods that assume a fixed trading unit, this agent dynamically adjusts the trade size based on account exposure, memory information, and risk constraints. Its decision-making process incorporates risk-aware references such as the scaled Kelly Criterion~\cite{Kelly_criterion} and conditional value-at-risk (CVaR~\cite{CVaR}) estimates to maintain sensitivity to downside risk under uncertainty. Ultimately, these two agents work together to simultaneously output trade direction and position size, achieving explicit control over capital risk and potential drawdown.


\begin{table*}[t]
\centering
\caption{Performance comparison of different models on five stocks}
\label{tab:model_comparison}
\resizebox{\textwidth}{!}{%
\begin{tabular}{l S S S S S S S S S S S S S S S S}
\Xhline{2\arrayrulewidth}
\multirow{2}{*}{\textbf{Models}} & \multicolumn{3}{c}{\textbf{TSLA}} & \multicolumn{3}{c}{\textbf{AAPL}} & \multicolumn{3}{c}{\textbf{AMZN}} & \multicolumn{3}{c}{\textbf{NFLX}} & \multicolumn{3}{c}{\textbf{COIN}} \\
\cmidrule(lr){2-4} \cmidrule(lr){5-7} \cmidrule(lr){8-10} \cmidrule(lr){11-13} \cmidrule(lr){14-16}
 & CR\%↑ & SR↑ & {MDD\%↓} & CR\%↑ & SR↑ & {MDD\%↓} & CR\%↑ & SR↑ & {MDD\%↓} & CR\%↑ & SR↑ & {MDD\%↓} & CR\%↑ & SR↑ & {MDD\%↓} \\
\Xhline{2\arrayrulewidth}
\rowcolor{myred!10}
\multicolumn{16}{c}{\textit{Market Baseline}} \\ \Xhline{2\arrayrulewidth}
Random    & -32.13 & -0.91 & 62.10 & -34.21 & -0.28 & 59.40 & -52.35 & -0.32 & 61.20 & -1.04 & -0.02 & 22.60 & -0.22 & -0.01 & 17.60 \\
Buy and Hold~\cite{b&h}    & -16.40 & -0.71 & 63.00 & -16.03 & -0.54 & 47.94 & -15.53 & -0.45 & 68.91 & 6.84 & 0.42 & 23.08 & -7.65 & -0.01 & 16.73 \\
\Xhline{2\arrayrulewidth}
\rowcolor{myyellow!30}
\multicolumn{16}{c}{\textit{Rule-Based Methods}} \\ \Xhline{2\arrayrulewidth}
MACD~\cite{macd}   & -49.21 & -0.63 & 74.20 & -46.83 & -0.57 & 61.40 & -44.92 & -0.49 & 70.30 & -7.42  & -0.18 & 27.90 & -9.87  & -0.21 & 19.80  \\
RSI~\cite{rsi}    & -46.13 & -0.59 & 72.80 & -44.28 & -0.54 & 59.70 & -42.11 & -0.46 & 68.20 & -6.98  & -0.16 & 26.70 & -8.65  & -0.19 & 19.30  \\
\Xhline{2\arrayrulewidth}
\rowcolor{mygreen!50}
\multicolumn{16}{c}{\textit{Reinforcement Learning Methods
}} \\ \Xhline{2\arrayrulewidth}
A3C~\cite{FinRL_a3c}    & -92.71 & -1.28 & 96.60 & -81.55 & -1.14 & 73.40 & -58.24 & -0.61 & 72.80 & -5.24  & -0.12 & 23.60 & -6.10  & -0.09 & 18.60  \\
DQN~\cite{FinRL_dqn}    & -78.25 & -0.77 & 80.62 & -72.48 & -0.92 & 70.30 & -52.77 & -0.48 & 68.10 & 0.78  & 0.08 & 22.10 & -4.25  & -0.06 & 18.20  \\
PPO~\cite{FinRL_ppo}    & -73.76 & -0.56 & 69.88 & -58.33 & -0.55 & 61.40 & -43.12 & -0.39 & 64.90 & 1.72  & 0.02 & 21.30 & -2.04  & -0.04 & 17.80 \\
\Xhline{2\arrayrulewidth}
\rowcolor{myblue}
\multicolumn{16}{c}{\textit{LLM-Based Agent
}} \\ \Xhline{2\arrayrulewidth}
FinGPT~\cite{FinGPT}    & -95.19 & -1.07 & 94.36 & -85.22 & -1.52 & 74.60 & -56.87 & -0.54 & 70.30 & 2.01  & 0.01  & 22.30 & -3.63 & -0.04 & 17.00 \\
FinAgent~\cite{finagent}  & -74.31 & -0.89 & 85.65 & -78.00 & -1.09 & 71.50 & -49.64 & -0.53 & 61.20 & 1.12  & 0.02  & 21.50 & -2.10 & -0.03 & 17.50 \\
FINMEM~\cite{finmem}    & -44.03 & -0.52 & 72.10 & -45.88 & -0.54 & 47.80 & -41.21 & -0.36 & 66.10 & 0.41  & 0.05  & 25.20 & 0.13  & 0.01  & 19.20 \\
FINCON~\cite{fincon}    & 7.76  & 0.38  & 59.13 & -16.02 & -0.13 & 29.03 & -3.30  & -0.02 & 54.77 & 7.92  & 0.63  & 20.70 & 7.35  & \textbf{1.03}  & 15.10 \\
\Xhline{2\arrayrulewidth}
\rowcolor{mypurple!20}
\textbf{FinRS}    & \textbf{54.99} & \textbf{0.67} & \textbf{42.34} & \textbf{60.28} & \textbf{0.69} & \textbf{18.44} & \textbf{52.10} & \textbf{0.64} & \textbf{36.92} & \textbf{16.85} & \textbf{0.87} & \textbf{20.05} & \textbf{14.74} & 0.92 & \textbf{14.05} \\
\Xhline{2\arrayrulewidth}
\end{tabular}%
}
\end{table*}

\subsection{Multi-Scale Reward Reflection}


To promote stable and risk-sensitive trading, this module integrates short- and long-term feedback. At each timestep, the agent's action is evaluated against a sliding-window account P\&L benchmark, producing structured rewards that reflect both potential returns and potential downside. The agent then performs self-reflection to adjust its actions. Unlike approaches that rely solely on next-day price changes, our mechanism incorporates price trends across multiple timescales, mitigating myopic responses and enhancing sensitivity to market volatility.

To embed risk-awareness, we design a trend-aware reward function leveraging multi-scale market momentum signals. Specifically, for each timestep $t$, we compute three future price trends:
$M_{t}^{s} = price[t+1] - price[t]$: 1-day trend (short-term),
$M_{t}^{m} = price[t+7] - price[t]$: 7-day trend (mid-term),
$M_{t}^{l} = price[t+30] - price[t]$: 30-day trend (long-term).
We define the multi-timescale momentum score $M_t$ as:

\begin{equation}
    M_t = M_{t}^{s} + M_{t}^{m} + M_{t}^{l}
\end{equation}
This score represents the aggregated expected trend across multiple timescales (e.g., 1-day, 7-day, and 30-day). The reward at time $t$, $Reward_{t}$ is defined as follows:

\begin{equation}
Reward_{t} = \left\{\begin{matrix}
-(M_{t})^{2}, & position_{t} =position_{t-1} \\  
position_{t} \times M_{t}, & otherwise \\
\end{matrix}\right.
\end{equation}

\begin{equation}
    position_{t} = position_{t - 1} + d_t \times q_t 
\end{equation}
where $d_t$ and $q_t$ denote the direction and quantity of the trading decision.

This reward design serves two risk-sensitive objectives:
(\rmnum{1}) Trend-aligned risk-adjusted returns: The agent is positively reinforced when its decision aligns with future market movement, scaled across horizons to balance short-term volatility against long-term growth potential.
(\rmnum{2}) Penalty for risk of inertia: During high-volatility periods, passivity is especially costly, as it amplifies exposure to missed opportunities. To address this, we introduce a nonlinear penalty term that grows quadratically with market fluctuations when the agent keeps positions unchanged, discouraging passivity and promoting proactive yet risk-aware actions.

\section{Experiments}

\subsection{Experimental Setup}
\textbf{(i) Datasets.} We use real-world financial data from authoritative sources: (1) daily open-high-low-close-volume (OHLCV) prices from Yahoo Finance; (2) company and macro news from Finnhub; and (3) 10-Q/10-K filings~\cite{financial_reporting} from the SEC EDGAR API. All data are processed into daily time series.  
\textbf{(ii) Baselines.} We compare FinRS with four LLM-based agents (FinGPT~\cite{FinGPT}, FinMem~\cite{finmem}, FinAgent~\cite{finagent}, FinCon~\cite{fincon}), three DRL~\cite{DRL} methods (A2C, PPO, DQN), two rule-based strategies (MACD, RSI), and two market baselines (Random, Buy-and-Hold).
\textbf{(iii) Metrics.} We report Cumulative Return~\cite{cr} (CR\%), Sharpe Ratio~\cite{sr} (SR), and Maximum Drawdown (MDD\%), which respectively reflect profitability, risk-adjusted efficiency, and downside risk. 
\textbf{(iv) Implementation.} All LLM agents run with GPT-4o~\cite{gpt4O} (temperature 0.7). Training covered Jan 2024–Feb 2025; testing spanned Mar 2025–Apr 2025, which includes the U.S. election and other significant macro events.

\subsection{Main Results}
We evaluate FinRS on five representative stocks (TSLA, AAPL, AMZN, NFLX, COIN), covering diverse sectors and volatility levels. Table~\ref{tab:model_comparison} shows that FinRS consistently outperforms all baselines, achieving higher CR, superior SR, and lower MDD across the board.  

As shown in Table~\ref{tab:model_comparison}, the full FinRS configuration achieves the best performance across all five stocks. In particular, FinRS demonstrates strong robustness on stocks that experienced high volatility during the test period, such as TSLA, AAPL, and AMZN. Other LLM agents and DRL methods occasionally suffered large losses, generated negative returns, or failed to outperform simpler benchmark strategies during turbulent market periods. In contrast, FinRS consistently maintained cumulative returns above 50\% while effectively controlling drawdowns. For relatively stable stocks like NFLX and COIN, FinRS achieved returns more than twice those of competing approaches. These results demonstrate that FinRS’s risk-sensitive design enhances profitability while maintaining stable trading behavior across diverse market conditions, showing resilience to volatility and uncertainty.


\begin{table*}[htbp]
\centering
\footnotesize
\caption{Ablation Results on TSLA, AAPL and AMZN}
\label{tab:ablation_single}
\resizebox{\textwidth}{!}{%
\begin{tabular}{c c c c | c c c | c c c | c c c}
\toprule
\textbf{RS} & \textbf{FIP} & \textbf{MN} & \textbf{MTR} & \multicolumn{3}{c|}{\textbf{TSLA}} & \multicolumn{3}{c|}{\textbf{AAPL}} &\multicolumn{3}{c}{\textbf{AMZN}} \\
           &              &             &             & CR\%↑ & SR↑ & MDD\%↓ & CR\%↑ & SR↑ & MDD\%↓ & CR\%↑ & SR↑ & MDD\%↓ \\
\midrule
           & \checkmark  & \checkmark  & \checkmark   & 39.19  & 0.45 & 82.89    & 45.96 & 0.49 & 36.10 & 36.98 & 0.43 & 73.71    \\
\checkmark  &              & \checkmark           & \checkmark  & 41.57     & 0.49 & 48.78      & 48.57     & 0.54 & 21.24  & 39.38 & 0.50 & 41.76    \\
\checkmark          & \checkmark            &             & \checkmark           & 45.41     & 0.53 & 55.40      & 52.78     & 0.60 & 24.13   & 43.67 & 0.55 & 49.53   \\
\checkmark          & \checkmark            & \checkmark           &             & 48.68     & 0.59 & 59.07      & 57.36     & 0.66 & 25.73  & 47.24 & 0.60 & 51.50    \\
\checkmark          & \checkmark            & \checkmark           & \checkmark           & \textbf{54.99} & \textbf{0.67} & \textbf{42.34} & \textbf{60.28} & \textbf{0.69} & \textbf{18.44} & \textbf{52.10} & \textbf{0.64} & \textbf{36.92} \\
\bottomrule
\end{tabular}
}
\end{table*}

\subsection{Ablation Studies}
We evaluated the contribution of each FinRS component through ablation experiments on three representative assets (TSLA, AAPL, AMZN). The results in Table~\ref{tab:ablation_single} show clear degradation once any module is removed.

\begin{itemize}[left=0.3cm]
\item \textbf{Risk-Sensitive Reasoning (RS)}:Embeds risk awareness by incorporating position information and volatility adjustment. Removing RS causes the sharpest decline: for TSLA, the CR drops from 54.99\% to 39.19\% (–15.8), while MDD sharply worsens from 42.34\% to 82.89\%. AAPL and AMZN also lose more than 14–15 points in CR. This confirms that RS is indispensable for both profitability and drawdown control.

\item \textbf{Financial Insight Prompting (FIP)}: A prompt strategy emphasizing causal chains, momentum, and probabilistic reasoning. Without FIP, TSLA’s CR decreases from 54.99\% to 41.57\%, and SR falls from 0.67 to 0.49. AAPL records an 11.7-point CR drop, suggesting that FIP enhances reasoning depth and contributes to improved returns.  

\item \textbf{Market News Integration (MN)}: Incorporates macroeconomic and geopolitical signals. Excluding MN reduces TSLA’s CR to 45.41\% and raises MDD from 42.34\% to 55.40\%. AAPL and AMZN show similar CR declines (–7.5 and –8.4), indicating that MN broadens the information scope and provides crucial contextual signals, especially in volatile markets.

\item \textbf{Multi-Timescale Reward (MTR)}: Adds a lightweight self-assessment after each decision. Removing MTR reduces TSLA’s CR to 48.68\% and worsens MDD by 16.7 points. AAPL also suffers higher drawdowns, suggesting that reflective feedback across horizons improves foresight and mitigates losses.
\end{itemize}

Overall, the full FinRS model achieves the best CR, SR, and MDD across all assets. The ablation results confirm that each module contributes complementary benefits, with RS being the most critical driver of robustness and profitability.

\section{Conclusion}

We introduced FinRS, a risk-sensitive LLM trading framework for multi-stage sequential decision-making in realistic markets. By integrating rolling-window analysis, hierarchical information processing, and a dual-decision system, FinRS aligns trades with both portfolio exposure and downside risks. Experiments show that this design improves prediction, reduces drawdowns, and enhances robustness under volatility. Our findings highlight the value of embedding risk-aware reasoning in LLM trading agents,  laying the foundation for more reliable and sustainable algorithmic trading.



\bibliographystyle{IEEEbib}
\bibliography{strings,refs}

@article{FinGPT,
  author       = {Hongyang Yang and
                  Xiao{-}Yang Liu and
                  Christina Dan Wang},
  title        = {FinGPT: Open-Source Financial Large Language Models},
  journal      = {CoRR},
  volume       = {abs/2306.06031},
  year         = {2023},
  doi          = {10.48550/ARXIV.2306.06031},
  eprinttype    = {arXiv},
  eprint       = {2306.06031},
}

@inproceedings{finmem,
  title={FinMem: A performance-enhanced LLM trading agent with layered memory and character design},
  author={Yu, Yangyang and Li, Haohang and Chen, Zhi and Jiang, Yuechen and Li, Yang and Zhang, Denghui and Liu, Rong and Suchow, Jordan W and Khashanah, Khaldoun},
  booktitle={Proceedings of the AAAI Symposium Series},
  volume={3},
  number={1},
  pages={595--597},
  year={2024}
}

@inproceedings{finagent,
  title={A multimodal foundation agent for financial trading: Tool-augmented, diversified, and generalist},
  author={Zhang, Wentao and Zhao, Lingxuan and Xia, Haochong and Sun, Shuo and Sun, Jiaze and Qin, Molei and Li, Xinyi and Zhao, Yuqing and Zhao, Yilei and Cai, Xinyu and others},
  booktitle={Proceedings of the 30th ACM SIGKDD Conference on Knowledge Discovery and Data Mining},
  pages={4314--4325},
  year={2024}
}

@inproceedings{finvision,
  title={FinVision: A Multi-Agent Framework for Stock Market Prediction},
  author={Fatemi, Sorouralsadat and Hu, Yuheng},
  booktitle={Proceedings of the 5th ACM International Conference on AI in Finance},
  pages={582--590},
  year={2024}
}

@article{fincon,
  title={Fincon: A synthesized llm multi-agent system with conceptual verbal reinforcement for enhanced financial decision making},
  author={Yu, Yangyang and Yao, Zhiyuan and Li, Haohang and Deng, Zhiyang and Jiang, Yuechen and Cao, Yupeng and Chen, Zhi and Suchow, Jordan and Cui, Zhenyu and Liu, Rong and others},
  journal={Advances in Neural Information Processing Systems},
  volume={37},
  pages={137010--137045},
  year={2024}
}

@article{tradingagents,
  title={TradingAgents: Multi-Agents LLM Financial Trading Framework},
  author={Xiao, Yijia and Sun, Edward and Luo, Di and Wang, Wei},
  journal={arXiv preprint arXiv:2412.20138},
  year={2024}
}

@incollection{Kelly_criterion,
  title={Portfolio choice and the Kelly criterion},
  author={Thorp, Edward O},
  booktitle={Stochastic optimization models in finance},
  pages={599--619},
  year={1975},
  publisher={Elsevier}
}

@article{CVaR,
  title={Optimization of conditional value-at-risk},
  author={Rockafellar, R Tyrrell and Uryasev, Stanislav and others},
  journal={Journal of risk},
  volume={2},
  pages={21--42},
  year={2000},
  publisher={Citeseer}
}

@article{gpt4O,
  title={Gpt-4o system card},
  author={Hurst, Aaron and Lerer, Adam and Goucher, Adam P and Perelman, Adam and Ramesh, Aditya and Clark, Aidan and Ostrow, AJ and Welihinda, Akila and Hayes, Alan and Radford, Alec and others},
  journal={arXiv preprint arXiv:2410.21276},
  year={2024}
}

@inproceedings{FinRL_ppo,
  title={Option-Driven Sentiment in FinRL: a PPO Approach to Trading},
  author={Li, Shenjian and Yu, Mingxuan and Dossor, Freddie},
  booktitle={2025 IEEE 11th International Conference on Intelligent Data and Security (IDS)},
  pages={62--64},
  year={2025},
  organization={IEEE Computer Society}
}

@article{FinRL_dqn,
  title={Improving financial trading decisions using deep Q-learning: Predicting the number of shares, action strategies, and transfer learning},
  author={Jeong, Gyeeun and Kim, Ha Young},
  journal={Expert Systems with Applications},
  volume={117},
  pages={125--138},
  year={2019},
  publisher={Elsevier}
}

@inproceedings{FinRL_a3c,
  title={An asynchronous advantage actor-critic reinforcement learning method for stock selection and portfolio management},
  author={Kang, Qinma and Zhou, Huizhuo and Kang, Yunfan},
  booktitle={Proceedings of the 2nd International Conference on Big Data Research},
  pages={141--145},
  year={2018}
}

@article{macd,
  title={Predicting stock price trend using MACD optimized by historical volatility},
  author={Wang, Jian and Kim, Junseok},
  journal={Mathematical Problems in Engineering},
  volume={2018},
  number={1},
  pages={9280590},
  year={2018},
  publisher={Wiley Online Library}
}

@article{rsi,
  title={Validity and reliability of the reflux symptom index (RSI)},
  author={Belafsky, Peter C and Postma, Gregory N and Koufman, James A},
  journal={Journal of voice},
  volume={16},
  number={2},
  pages={274--277},
  year={2002},
  publisher={Elsevier}
}

@article{b&h,
  title={Thou shalt buy and hold},
  author={Shiryaev, Albert and Xu, Zuoquan and Zhou, Xun Yu},
  journal={Quantitative finance},
  volume={8},
  number={8},
  pages={765--776},
  year={2008},
  publisher={Taylor \& Francis}
}

@article{llm,
  title={ChatGPT for good? On opportunities and challenges of large language models for education},
  author={Kasneci, Enkelejda and Se{\ss}ler, Kathrin and K{\"u}chemann, Stefan and Bannert, Maria and Dementieva, Daryna and Fischer, Frank and Gasser, Urs and Groh, Georg and G{\"u}nnemann, Stephan and H{\"u}llermeier, Eyke and others},
  journal={Learning and individual differences},
  volume={103},
  pages={102274},
  year={2023},
  publisher={Elsevier}
}

@article{mdd,
  title={Maximum drawdown},
  author={Magdon-Ismail, Malik and Atiya, Amir F},
  journal={Risk Magazine},
  volume={17},
  number={10},
  pages={99--102},
  year={2004}
}

@article{sr,
  title={The sharpe ratio},
  author={Sharpe, William F},
  journal={Streetwise--the Best of the Journal of Portfolio Management},
  volume={3},
  number={3},
  pages={169--85},
  year={1998},
  publisher={Princeton University Press NJ}
}

@article{cr,
  title={A monthly effect in stock returns},
  author={Ariel, Robert A},
  journal={Journal of financial economics},
  volume={18},
  number={1},
  pages={161--174},
  year={1987},
  publisher={Elsevier}
}

@article{agent,
  title={Agent-based models of financial markets},
  author={Samanidou, Egle and Zschischang, Elmar and Stauffer, Dietrich and Lux, Thomas},
  journal={Reports on Progress in Physics},
  volume={70},
  number={3},
  pages={409},
  year={2007},
  publisher={IOP Publishing}
}

@article{DRL,
  title={Deep reinforcement learning for trading—A critical survey},
  author={Millea, Adrian},
  journal={Data},
  volume={6},
  number={11},
  pages={119},
  year={2021},
  publisher={MDPI}
}

@article{financial_reporting,
  title={Automating financial reporting with natural language processing: A review and case analysis},
  author={Oyewole, Adedoyin Tolulope and Adeoye, Omotayo Bukola and Addy, Wilhelmina Afua and Okoye, Chinwe Chinazo and Ofodile, Onyeka Chrisanctus and Ugochukwu, Chinonye Esther},
  journal={World Journal of Advanced Research and Reviews},
  volume={21},
  number={3},
  pages={575--589},
  year={2024},
  publisher={World Journal of Advanced Research and Reviews}
}

@article{llm_reasoning,
  title={A survey of frontiers in llm reasoning: Inference scaling, learning to reason, and agentic systems},
  author={Ke, Zixuan and Jiao, Fangkai and Ming, Yifei and Nguyen, Xuan-Phi and Xu, Austin and Long, Do Xuan and Li, Minzhi and Qin, Chengwei and Wang, Peifeng and Savarese, Silvio and others},
  journal={arXiv preprint arXiv:2504.09037},
  year={2025}
}

@inproceedings{llm_decision,
  title={Enhancing decision-making for llm agents via step-level q-value models},
  author={Zhai, Yuanzhao and Yang, Tingkai and Xu, Kele and Feng, Dawei and Yang, Cheng and Ding, Bo and Wang, Huaimin},
  booktitle={Proceedings of the AAAI Conference on Artificial Intelligence},
  volume={39},
  number={25},
  pages={27161--27169},
  year={2025}
}

@article{risk_control,
  title={Managed care and capitation in California: how do physicians at financial risk control their own utilization?},
  author={Kerr, Eve A and Mittman, Brian S and Hays, Ron D and Siu, Albert L and Leake, Barbara and Brook, Robert H},
  journal={Annals of Internal Medicine},
  volume={123},
  number={7},
  pages={500--504},
  year={1995},
  publisher={American College of Physicians}
}

@article{deepseek-v3,
  title={Deepseek-v3 technical report},
  author={Liu, Aixin and Feng, Bei and Xue, Bing and Wang, Bingxuan and Wu, Bochao and Lu, Chengda and Zhao, Chenggang and Deng, Chengqi and Zhang, Chenyu and Ruan, Chong and others},
  journal={arXiv preprint arXiv:2412.19437},
  year={2024}
}

\end{document}